\documentclass[letterpaper]{article} % DO NOT CHANGE THIS
\usepackage{aaai25}  % DO NOT CHANGE THIS
\usepackage{times}  % DO NOT CHANGE THIS
\usepackage{helvet}  % DO NOT CHANGE THIS
\usepackage{courier}  % DO NOT CHANGE THIS
\usepackage[hyphens]{url}  % DO NOT CHANGE THIS
\usepackage{graphicx} % DO NOT CHANGE THIS
\usepackage{natbib}  % DO NOT CHANGE THIS AND DO NOT ADD ANY OPTIONS TO IT
\usepackage{caption} % DO NOT CHANGE THIS AND DO NOT ADD ANY OPTIONS TO IT
\usepackage{algorithm}
\usepackage{algorithmic}

\usepackage{newfloat}
\usepackage{listings}

\usepackage{amsmath}
\usepackage{amssymb}
\usepackage{tabularx}
\usepackage{graphicx}
\usepackage{subcaption}

\DeclareCaptionStyle{ruled}{labelfont=normalfont,labelsep=colon,strut=off} % DO NOT CHANGE THIS
\floatstyle{ruled}
\newfloat{listing}{tb}{lst}{}
\floatname{listing}{Listing}
\title{Exploiting Fine-Grained Skip Behaviors for Micro-Video Recommendation}
\author{
    Sanghyuck Lee, Sangkeun Park, Jaesung Lee\thanks{Corresponding author.}
}
\affiliations{
    Department of Artificial Intelligence, Chung-Ang University, Seoul, Republic of Korea\\
    \{tkdgur658, psk492, curseor\}@cau.ac.kr
}

\usepackage{bibentry}
\begin{document}

\maketitle

\begin{abstract}
The growing trend of sharing short videos on social media platforms, where users capture and share moments from their daily lives, has led to an increase in research efforts focused on micro-video recommendations. However, conventional methods oversimplify the modeling of skip behavior, categorizing interactions solely as positive or negative based on whether skipping occurs. This study was motivated by the importance of the first few seconds of micro-videos, leading to a refinement of signals into three distinct categories: highly positive, less positive, and negative. Specifically, we classify skip interactions occurring within a short time as negatives, while those occurring after a delay are categorized as less positive. The proposed dual-level graph and hierarchical ranking loss are designed to effectively learn these fine-grained interactions. Our experiments demonstrated that the proposed method outperformed three conventional methods across eight evaluation measures on two public datasets.
\end{abstract}


\section{Introduction}
Micro-videos typically refer to self-generated video content that is usually less than three minutes long, covering a wide range of daily life aspects such as the latest news, funny clips, and sports highlights~\cite{ZHANG2024104014}. According to a report by Vidico, expenditures on micro-video advertisements are expected to reach approximately 100 billion dollars in 2024, while video content is projected to account for 82\% of global internet traffic by 2025~\cite{MV1}. Furthermore, recent policies by major companies, such as the TikTok creator fund policy and YouTube creator partnership program, underscore the continued global expansion of investment in micro-videos~\cite{MV2}. Given this overwhelming abundance, access to micro-videos is primarily driven by recommendation algorithms rather than self-searching~\cite{MV3}, and user satisfaction declines when platforms repeatedly display videos that do not align with their interests~\cite{GU2024111299}. As a result, the need for highly sophisticated recommendation systems that can effectively analyze user preferences and identify potentially interesting micro-videos in a personalized manner has become even more critical~\cite{lu2023multi}.

\begin{figure}[!t]
\centerline{\includegraphics[width=\columnwidth]{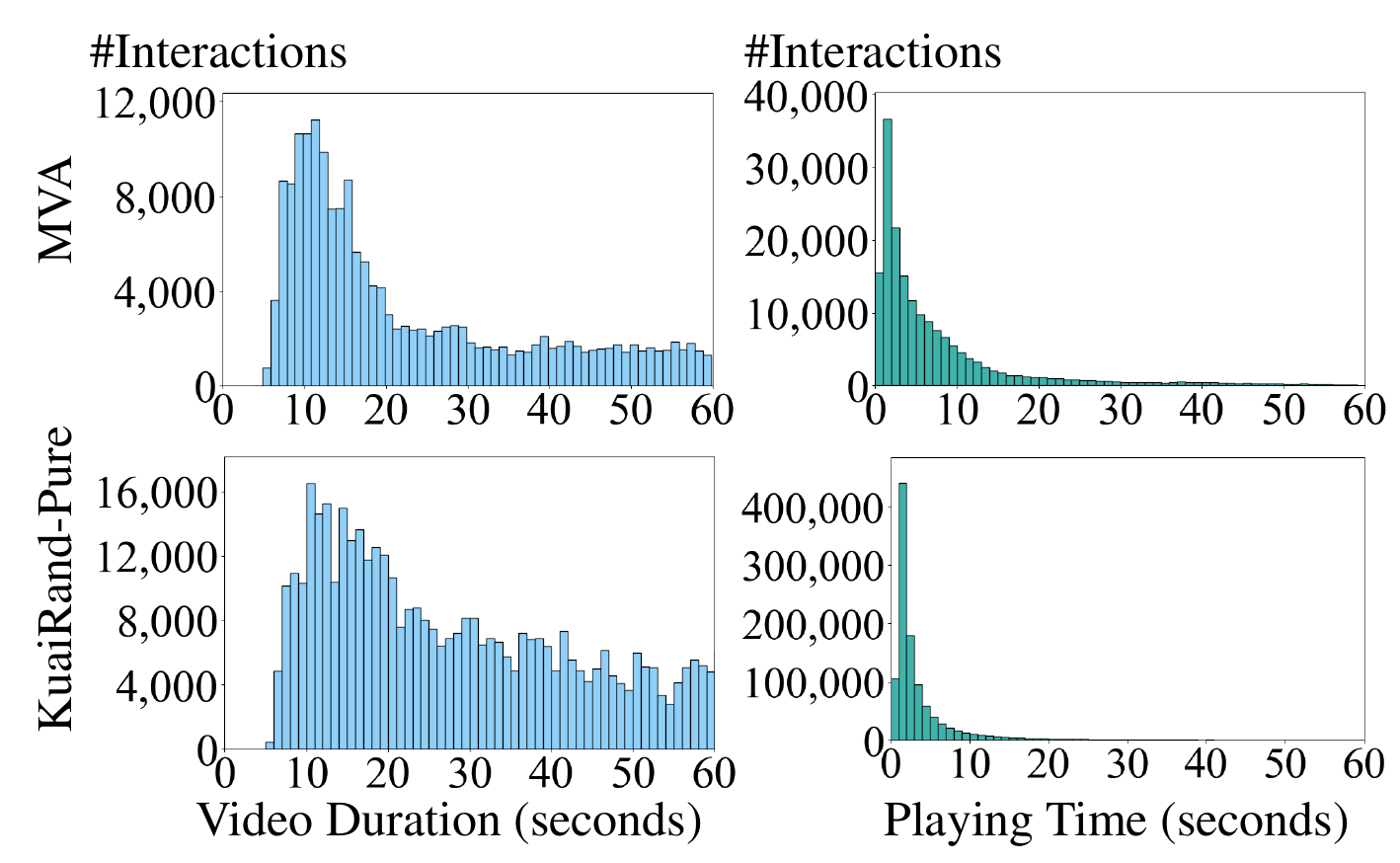}}
\vspace{-10pt} 
\caption{Distribution of video duration and playing time of the potent skipped interactions in two datasets, MVA~\cite{10.1145/3539618.3591713} and KuaiRand-Pure \cite{10.1145/3511808.3557624}. The figures include only interactions where the playing time is shorter than the video duration; thus, playing time can be considered indicative of the timing of skip behaviors. Most skips occur within the first five seconds of the video, while the distribution of video durations remains relatively uniform. The conventional approach~\cite{10.1145/3539618.3591713} based on playing time views incomplete viewing as negative, ignoring that users might form positive impressions early, causing slightly delayed skips. The histogram bin range has been truncated to 0-60 seconds for the sake of clarity.
}
\label{skip timing}
 \vspace{-16pt} 
\end{figure}

Conventional approaches for micro-video recommendation are roughly divided into two strategies. One strategy involves utilizing the multi-modal information of micro-videos~\cite{10.1145/3343031.3351034,10.1145/3340555.3355720}, and the other group aims to capture the different interests of users~\cite{10.1145/3394171.3413653,10.1145/3477495.3532081}. Despite their success, conventional methods have limited potential for further performance improvements due to their failure to effectively leverage the rich information that playing time conveys about both positive and negative interests, such as skip behavior~\cite{GU2024111299}. 

While browsing content or using the platform, users should watch for a few seconds before deciding whether to continue viewing the video or swipe down to move on to the next one~\cite{10.1145/3511808.3557220}. 
Due to this skippable nature of micro-videos, the significance of the first few seconds has been highlighted by multimedia stakeholders~\cite{Willis2024,Wang_2021,9377378,10.1007/978-3-031-05897-4_29}. These initial moments in micro-videos should be carefully considered to encourage users to watch the video until the end~\cite{MV4}. Similarly, in Figure~\ref{skip timing}, the statistics of Micro-video-A (MVA)~\cite{10.1145/3539618.3591713} and KuaiRand-Pure~\cite{10.1145/3511808.3557624} datasets show that most skips occurred within the first five seconds of interaction, suggesting a strong negative signal from users for the video content. In other words, a delayed skip may have different traits from this strong negative signal, which is our primary focus in this study.

Inspired by this motivation, we propose a dual-graph-based micro-video recommender, which contains a dual-level positive graph receiving help from the less positive interest separated from skip behavior interactions. Furthermore, negative interactions are integrated into the optimization process rather than being included in this graph construction, resulting in a hierarchical ranking loss.
We summarize the contributions of this paper as follows:
\begin{itemize}
    \item The proposed model, an adaptation of the conventional FRAME model that distinguishes between positive and negative interactions based on whether skipping occurs, refines the interaction types into three categories: highly positive, less positive, and negative. This improved approach demonstrates superior performance across eight evaluation measures on two datasets compared to the three conventional models.
    \item By considering the delayed skip as a less positive signal, the proposed dual-graphs using the dual-level positives demonstrate higher performance compared to both training with only the highly positive signal or training without distinguishing between the two levels.
    \item The quick skip is regarded as carrying a strong negative signal, which helps in improving training with conventional Bayesian Personalized Ranking (BPR) loss. 
\end{itemize}

\section{Related Work}

Early recommendations for micro-videos relied on standard collaborative filtering systems, which modeled interactions based solely on user and video identifiers (IDs). These approaches have since evolved to better capture the dynamic nature of user interests over time. For example, THACIL~\cite{chen2018temporal} utilized a hierarchical attention mechanism to capture video characteristics and user preferences, while UHMAN~\cite{liu2020user} recommended hashtags by analyzing video keywords within user histories. Later models, such as ALPINE~\cite{li2019routing} and MTIN~\cite{10.1145/3394171.3413653}, were designed to track and model user preferences across various time frames. DMR~\cite{lu2023multi} further introduced capabilities to capture both historical and predictive user interest trends.

User-item interactions have been shown to naturally form a bipartite graph, facilitating complex information extraction between nodes~\cite{wang2019neural,wang2020disentangled,rendle2020neural}. Given their inherent complexity, micro-videos necessitate analyses that incorporate visual, acoustic, and textual characteristics. Recommendations based on graph convolution networks (GCNs) typically integrate user-item interactions with multi-modal data. MMGCN~\cite{10.1145/3343031.3351034} exploits user-video bipartite graphs for each modality, enhancing user profiles through data aggregation from multi-hop neighboring nodes. DualGNN~\cite{wang2021dualgnn} introduced a preference learning module to fine-tune interest assessments across modalities, while ElimRec~\cite{liu2022elimrec} applied causal inference to minimize biases associated with single-modality focus. Following models, such as HUIGN~\cite{wei2021hierarchical} and HGCL~\cite{cai2022heterogeneous}, employed contrastive learning for hierarchical and heterogeneous understanding of user-video relationships, with A2BM2GL~\cite{cai2022adaptive} and LUDP~\cite{lei2023learning} further refining these approaches by optimizing graph weights and user preference modeling. GRCN~\cite{wei2020graph}, CONDE~\cite{liu2021concept}, and HHFAN~\cite{cai2021heterogeneous} have investigated techniques, such as graph refinement and subgraph construction to enhance computational efficiency and recommendation accuracy. In a different approach, FRAME~\cite{10.1145/3539618.3591713} proposed a refined recommendation method that utilizes dual-graph construction with video clips labeled by user skip behaviors.

To address the limitations of supervised learning for interaction modeling, different strategies have been investigated. SLMRec~\cite{tao2022self} improved the representation of feature patterns using data augmentation and contrastive learning in different modalities. Similarly, MMGCL~\cite{yi2022multi} applied data augmentation but introduced negative sampling techniques to enhance the learning of modality contribution and correlation.  MMSSL~\cite{wei2023multi} employed adversarially trained transformed instances for cross-modal semantic similarity-based contrastive learning, enhancing model generalization and interaction capture. InvRL~\cite{du2022invariant} addressed biased correlations from diverse data usage by clustering user-video interactions into different environments, learning invariant representations in each to model user preferences with causal insight.

Conventional GCNs inherently treat all neighbors equally during information aggregation, which is a drawback, as they assign the same weight to all interactions. To address this issue, attention mechanisms-based GCNs have emerged. UVCAN~\cite{liu2019user} independently embedded user histories and video features, modeling their dynamic interactions through a co-attention mechanism. MGAT~\cite{tao2020mgat} built a bipartite graph based on interactions and aggregated weights through a modality-specific attention mechanism to discern user modal preferences. MMKGV~\cite{liu2022multi} also employed an attention mechanism, constructing a knowledge graph based on video similarities to weigh user interactions. LightGT~\cite{wei2023lightgt} utilized a transformer-based self-attention block to capture complex patterns and interactions between user-video nodes, effectively using layers.

\begin{figure*}[!t]
\centerline{\includegraphics[width=\textwidth]{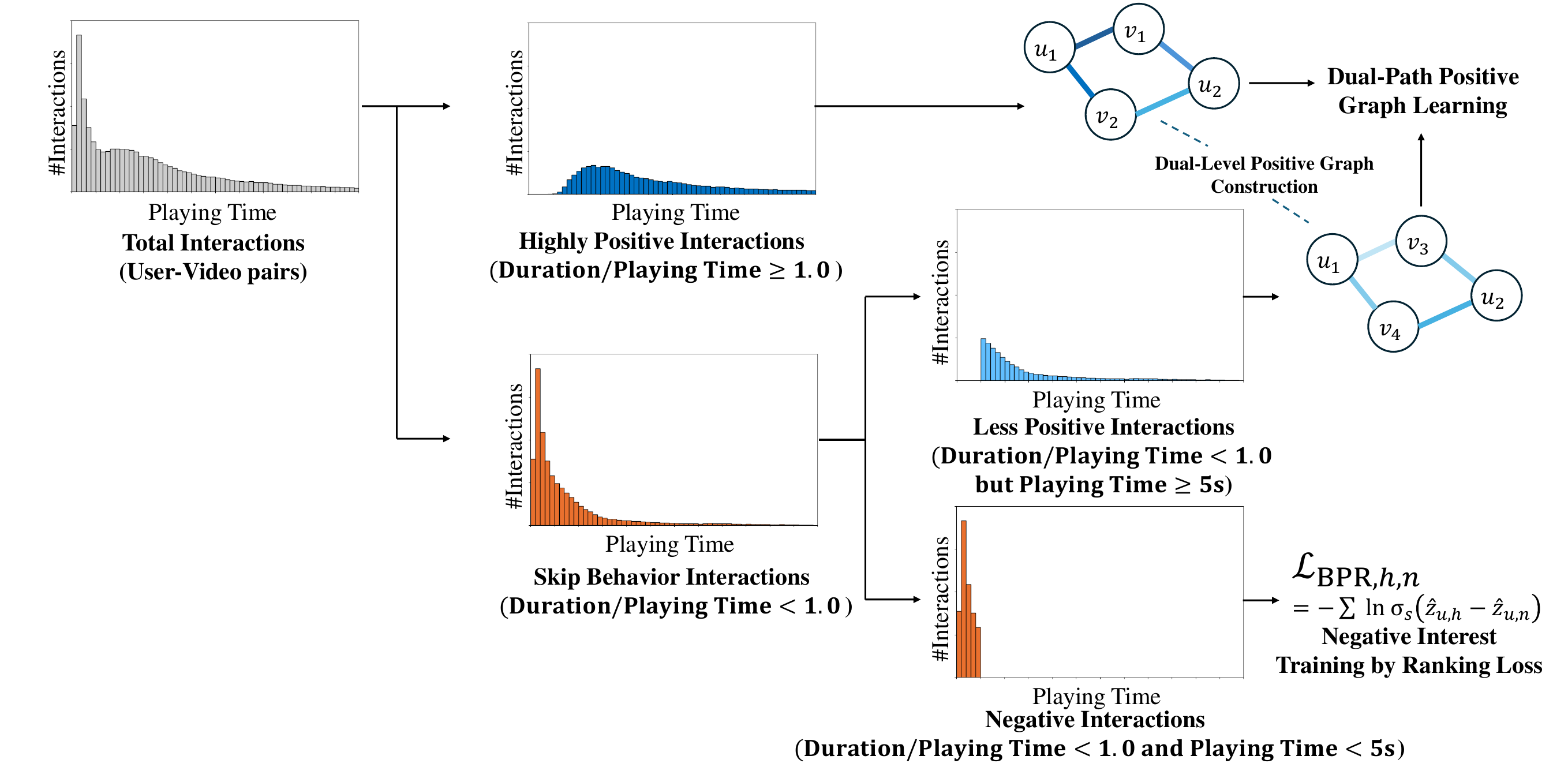}}
\caption{\label{fig:Interaction_Split} Dual-level positive graph construction and negative interest training by ranking loss. The total interactions are initially divided into Highly Positive Interactions and Skip Behavior Interactions based on Duration/Playing Time. Then, Skip Behavior Interactions are further divided into Less Positive Interactions and Negative Interactions, with Playing Time 5s as the threshold. Less Positive Interactions indicate a preference to continue watching the video beyond the initial 5 seconds, a period where skips are most frequent. Highly Positive Interactions and Less Positive Interactions each form individual adjacency graphs. These two adjacency graphs are utilized in dual-path positive graph learning. Negative Interactions help the model learn preference differences between interactions through a ranking loss.}
\end{figure*}

\section{Method}

\subsection{Problem Definition}

Let $\mathcal{U}=\left\{u_1, u_2, \cdots, u_{|\mathcal{U}|}\right\}$ and $\mathcal{V}=\left\{v_1, v_2, \cdots, v_{|\mathcal{V}|}\right\}$ denote the user set and the video set, respectively. The total historical interactions between users and videos are formatted as a sequence of triplets, $\mathcal{I} =\left\langle \left(u_i, v_j, y_k\right)_{k=1}^{|\mathcal{I}|} \right\rangle$ where $y_k$ represents the corresponding label indicating whether the video in the $k$-th interaction was 100\% viewed or skipped by the user in the $k$-th interaction. The skip case is denoted as zero, and the fully viewed case is denoted as one. For example, $\left(u_1, v_3, 1\right)$ means that the user $u_1$ watched all of the video $v_3$ and $\left(u_2, v_5, 0\right)$ means that the user $u_2$ skip the video $v_5$. The problem in this study can be formulated as follows.

\begin{figure*}[!t]
\centerline{\includegraphics[width=\textwidth]{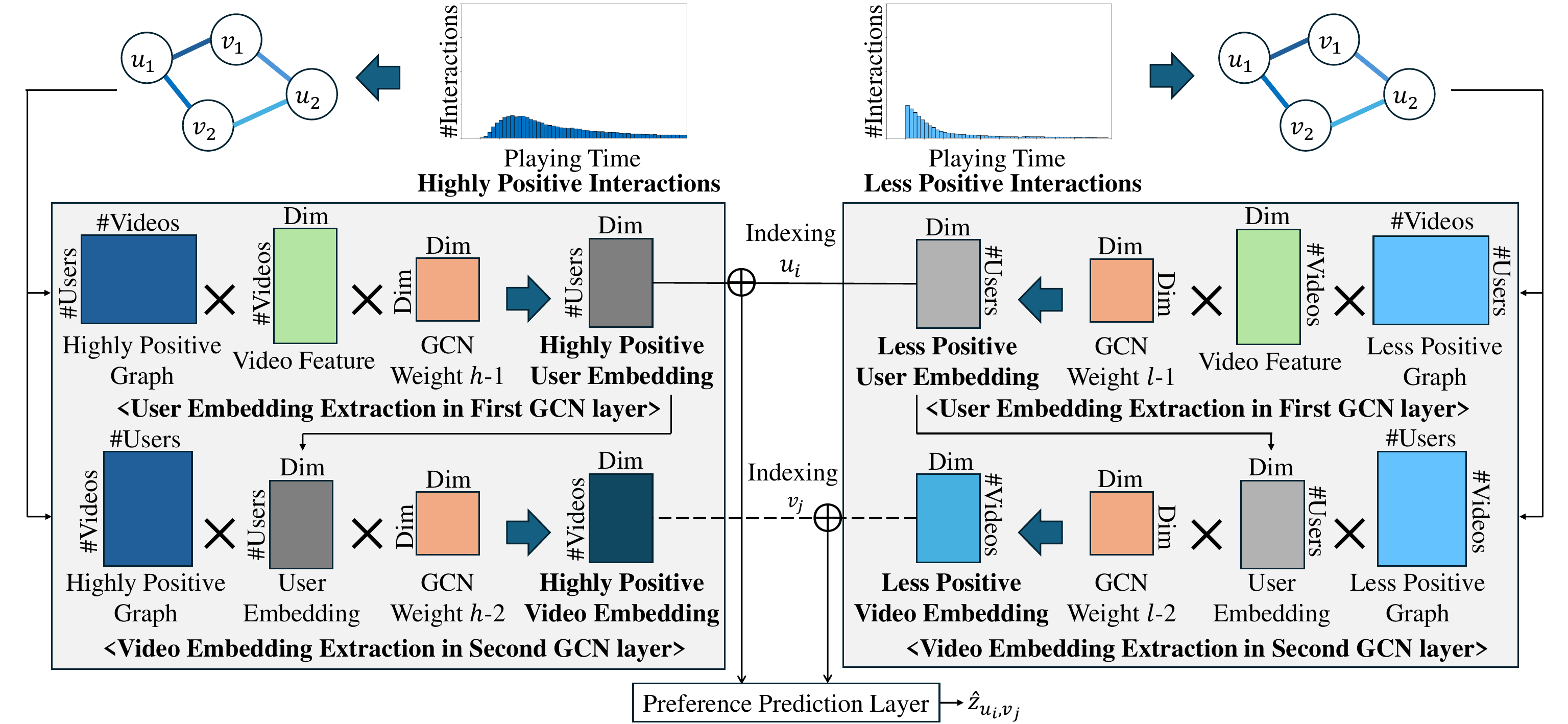}}
\caption{A schematic overview of the proposed dual-path positive graph learning. The video features are processed through distinct paths corresponding to the adjacency matrices from the highly and less positive graphs, reaching the preference prediction layer. The user embedding and video embedding generated from the two paths are then mean-pooled and concatenated. The fused features are passed through the prediction layer to output the preference score of the user $u_i$ for the video $v_j$.}
\label{Proposed}
\end{figure*}

\noindent \textbf{Input:} The total interactions $\mathcal{I}$; The visual features $X_v$ extracted from the image pixels of video frames by pre-trained image feature extraction model for all videos $v$ in $\mathcal{V}$.

\noindent \textbf{Output:} A micro-video recommendation model that estimates the preference score of a user $u$ given video $v$.

\subsection{Proposed Method}

\subsubsection{Dual-Level Positive Graph Construction.}

To exploit less positive interest from users, we construct two separate user-item interaction graphs by leveraging the skip behaviors observed in the first few seconds of video playback. These graphs are designed to capture highly and less positive signals from user interactions, respectively.

Given the user \(u\) in the set of users as \(\mathcal{U}\), the number of videos skipped within the first few seconds is denoted as \(N^{(u)}_l\), while the remaining videos in the interactions are denoted as \(N^{(u)}_h\). Duplicated videos caused by duplicate interaction are considered highly positive.

For the highly positive signal, we collect all videos from the interactions of the user \(u\) that were not skipped within the first few seconds to form the highly positive video set \(\mathcal{V}^h_{u}\), which is defined as
\begin{equation}
\mathcal{V}^h_{u} = \{v^{h,1}_{u}, v^{h,2}_{u}, \dots, v^{h,|\mathcal{V}^h_{u}|}_{u}\},
\end{equation}
where each \(v^{h,k}_{u}\) represents a video from the interaction history of user \(u\) that was not skipped in the first few seconds.

For the less positive signal, we focus on the videos during which the user \(u\) exhibited delayed skipping behavior after the first few seconds. This forms the less positive video set \(\mathcal{V}^l_{u}\), defined as
\begin{equation}
\mathcal{V}^l_{u} = \{v^{l,1}_{u}, v^{l,2}_{u}, \dots, v^{l,|\mathcal{V}^l_{u}|}_{u}\},
\end{equation}
where each \(v^{l,k}_{u}\) represents a video from the interaction history of user \(u\) that was skipped after the first few seconds.

Given the video sets with highly positive signal \(\mathcal{V}^h_{u}\) and less positive signal \(\mathcal{V}^l_{u}\) for each user, we construct two corresponding user-video interaction graphs. These graphs can be represented using the interaction matrices \(R^h\) and \(R^l\) for the highly and less positive relationships. For a set of users \(\mathcal{U}\) and videos \(\mathcal{V}\), the interaction matrices \(R^h\) and \(R^l\) are defined as
\begin{equation}
R^h_{ij} =
\begin{cases} 
1 & \text{if } v_j \in \mathcal{V}^h_{u_i}; \\
0 & \text{otherwise};
\end{cases}
\end{equation}
whereas
\begin{equation}
R^l_{ij} =
\begin{cases} 
1 & \text{if } v_j \in \mathcal{V}^l_{u_i}; \\
0 & \text{otherwise}.
\end{cases}
\end{equation}

Using these interaction matrices, we construct the dual-side adjacency matrices for the user-video graphs as
\begin{equation}
A^h = \begin{pmatrix}
0 & R^h \\
(R^h)^\top & 0
\end{pmatrix}, 
\end{equation}
whereas
\begin{equation}
A^l = \begin{pmatrix}
0 & R^l \\
(R^l)^\top & 0
\end{pmatrix}.
\end{equation}

These matrices are then normalized to construct the adjacency matrices \(\tilde{A}^h\) and \(\tilde{A}^l\) for the highly positive and less positive interactions, respectively. A symmetric normalization approach was used, where values were divided by the square root of the column and row degrees~\cite{10.1145/3397271.3401063}. The normalized adjacency matrices are given by
\begin{align}
\tilde{A}^h = (D^h)^{-\frac{1}{2}} A^h (D^h)^{-\frac{1}{2}}, 
\end{align}
whereas
\begin{align}
\tilde{A}^l = (D^l)^{-\frac{1}{2}} A^l (D^l)^{-\frac{1}{2}},
\end{align}
where \(D^h\) and \(D^l\) are diagonal matrices representing the degree of nodes in the highly and less positive graphs, respectively. Each entry \(D_{ii}\) in these matrices denotes the number of non-zero entries in the \(i\)-th row of the corresponding adjacency matrix.

By constructing and normalizing these graphs, we effectively separate highly and less positive signals, where less positive signals in this study were treated as negative signals in conventional studies, based on user skip behaviors which will enable the recommendation model to better understand and predict user preferences for micro-video content.

\subsubsection{Dual-Path Positive Graph Learning.}

Inspired by GCNs success in modeling higher-order interactions by aggregating information from different-hop neighbors, we extend this approach to user-video interactions. We derive user embeddings via the mechanism of embedding propagation in GCN models, where user embeddings are generated by aggregating the embeddings of their neighbors. For a user, neighbors are videos they have interacted with, while for a video, neighbors are users who interacted with it. 

We compute two sets of embeddings for each user, corresponding to highly and less positive interactions, as
\begin{align}
H_{u}^{h} &= \sigma \left( \left( \tilde{R}^{h} \right)^\top H_{v}^{(0)} W_{h}^{(1)} \right),
\end{align}
whereas
\begin{align}
H_{u}^{l} &= \sigma \left( \left( \tilde{R}^{l} \right)^\top H_{v}^{(0)} W_{l}^{(1)} \right),
\end{align}
where \( H_{v}^{(0)} \in \mathbb{R}^{|\mathcal{V}| \times d} \) denotes the initial video embedding matrix, which is derived from the visual features of the videos. The matrices \( \tilde{R}^{h} \) and \( \tilde{R}^{l} \) are the interaction matrices defined in Equations~(4) and~(5), respectively, and \( \sigma(\cdot) \) represents the nonlinear activation function. The matrices \( W_{h}^{(1)} \) and \( W_{l}^{(1)} \in \mathbb{R}^{d \times d} \) are trainable weight matrices. Thus, \( H_{u}^{h} \) and \( H_{u}^{l} \) are the user embeddings that capture highly and less positive interactions, respectively, and will be used for subsequent prediction tasks.

To capture higher-order relationships between videos, we perform a two-hop embedding propagation, which is formulated as
\begin{equation}
H_{v}^{h} = \sigma \left( \left( \tilde{R}^{h} \right)^\top H_{u}^{h} W_{h}^{(2)} \right),
\end{equation}
whereas
\begin{equation}
H_{v}^{l} = \sigma \left( \left( \tilde{R}^{l} \right)^\top H_{u}^{l} W_{l}^{(2)} \right),
\end{equation}
where $W_{h}^{(2)}$ and $W_{l}^{(2)} \in \mathbb{R}^{d \times d}$
are trainable weight matrices in the second GCN layer. Since both highly and less positive interactions represent multiple levels of user preference, we can combine these embeddings rather than treating them separately at this stage. Thus, we apply mean pooling to obtain the final user and video embedding representations, defined as
\begin{equation}
H_{u} = \text{Mean} \left( H_{u}^{h}, H_{u}^{l} \right),
\end{equation}
whereas
\begin{equation}
H_{v} = \text{Mean} \left( H_{v}^{h}, H_{v}^{l} \right).
\end{equation}

In summary, we obtain \( H_u \in \mathbb{R}^{|\mathcal{U}| \times d} \) and \( H_v \in \mathbb{R}^{|\mathcal{V}| \times d} \) as the embeddings of users and videos, respectively. These embeddings incorporate both highly and less positive interactions, allowing them to represent the users and videos collectively and effectively.

\subsection{Preference Prediction Layer}

After fusing dual-side interest embeddings of users and embeddings of videos from different GCN layers, we can now make the prediction. For a given user \( u_i \) and video \( v_j \), we index the corresponding user embedding \( \mathbf{h}_{u_i} \in \mathbb{R}^{d} \) and video embedding \( \mathbf{h}_{v_j} \in \mathbb{R}^{d} \) from \( H_{u} \) and \( H_{v} \), respectively. These embeddings are then concatenated and multiplied by a weight matrix, followed by a non-linear activation function, and finally multiplied by another weight matrix to output the final preference score
\begin{equation}
\hat{z}_{{u_i},{v_j}} = W^{(2)} \cdot \sigma (W^{(1)} \cdot [\mathbf{h}_{u_i}, \mathbf{h}_{v_j}]),
\end{equation}
where \( W^{(1)} \in \mathbb{R}^{2d \times d} \) and \( W^{(2)} \in \mathbb{R}^{d \times 1} \) are the weight matrices, and \([ \cdot , \cdot ]\) denotes the concatenation operation.

\subsection{Optimization}

\subsubsection{BPR Loss with Highly/Less Positive and Negative Samples.}

In this study, we extend the traditional BPR loss by incorporating highly positive (\( s \)), less positive (\( w \)), and negative (\( n \)) samples. For each training interaction involving a user \( u \) and a video $v$, we construct triplets by sampling two additional items to get one each of highly/less positive and negative. This triplet-based sampling strategy ensures that the model learns both from strong preference interest and from comparisons between varying levels of user preference and non-preference.

The BPR loss is computed twice: once using the highly and less positive items and once using the highly positive and negative items. The final BPR loss is the average of these two, which helps in balancing the ability of the model to rank items within different levels of user preference. The BPR loss for the highly and less positive interactions is defined as
\begin{equation}
\mathcal{L}_{\text{BPR}, h,l} = -\sum_{(u, h, l) \in D_{h,l}} \ln \sigma_s(\hat{z}_{u,h} - \hat{z}_{u,l}),
\end{equation}
where $\sigma_s$ is the sigmoid function, \( \hat{z}_{u,h} \) and \( \hat{z}_{u,l} \) are the predicted scores for the highly and less positive items, respectively, for user \( u \), and \( D_{h,l} \) represents the set of all training triples \( (u, h, l) \). This step ensures that the model can effectively rank items, even among those that the user has already shown some preference for, refining the precision of the recommendation system. Similarly, the BPR loss for the highly positive and negative items is defined as
\begin{equation}
\mathcal{L}_{\text{BPR},h,n} = -\sum_{(u, h, n) \in D_{h,n}} \ln \sigma_s(\hat{z}_{u,h} - \hat{z}_{u,n}),
\end{equation}
where \( \hat{z}_{u,h} \) and \( \hat{z}_{u,n} \) are the predicted scores for the highly positive and negative items, respectively, for user \( u \), and \( D_{h,n} \) represents the set of all training triples \( (u, h, n) \). This step enhances the ability of the model to distinguish between items that are highly preferred and those that are not preferred at all, enhancing the discrimination power of the model. The overall BPR loss is then computed as the average of these two losses
\begin{equation}
\mathcal{L}_{\text{BPR}} = \frac{1}{2} \left( \mathcal{L}_{\text{BPR}, h,l} + \mathcal{L}_{\text{BPR}, h,n} \right).
\end{equation}

Averaging these losses ensures that the model maintains a balanced perspective, optimizing both intra-preference ranking between highly and less positive items and inter-preference ranking between highly positive and negative items. 

\begin{table}[!t]
\begin{tabular}{cccc}
\hline
Dataset                                                                  & \#Users & \#Videos & \#Interactions \\ \hline
Micro-video-A                                                            & 12,739  & 58,291   & 342,694        \\
\begin{tabular}[c]{@{}c@{}}KuaiRand-Pure \\ (Random policy)\end{tabular} & 27,285  & 7,583    & 1,186,059      \\ \hline
\end{tabular}
\caption{\label{tab:dataset_statistics} Brief statistics of datasets employed in our study}
\end{table}

\begin{table*}[!t]
\centering
\begin{tabular}{c cccccccc}
\hline
Model & Precision@3 & Recall@3 & MAP@3 & NDCG@3 & Precision@5 & Recall@5 & MAP@5 & NDCG@5 \\ 
\hline
Proposed & \textbf{\begin{tabular}[c]{@{}c@{}}0.573*\\ ($\pm$ 0.002)\end{tabular}} & \textbf{\begin{tabular}[c]{@{}c@{}}0.623*\\ ($\pm$ 0.002)\end{tabular}} & \textbf{\begin{tabular}[c]{@{}c@{}}0.739*\\ ($\pm$ 0.002)\end{tabular}} & \textbf{\begin{tabular}[c]{@{}c@{}}0.790*\\ ($\pm$ 0.002)\end{tabular}} & \textbf{\begin{tabular}[c]{@{}c@{}}0.540*\\ ($\pm$ 0.002)\end{tabular}} & \textbf{\begin{tabular}[c]{@{}c@{}}0.882* \\ ($\pm$ 0.001)\end{tabular}} & \textbf{\begin{tabular}[c]{@{}c@{}}0.731*\\ ($\pm$ 0.002)\end{tabular}} & \textbf{\begin{tabular}[c]{@{}c@{}}0.812*\\ ($\pm$ 0.001)\end{tabular}} \\
BM3               & \begin{tabular}[c]{@{}c@{}}0.546 \\ ($\pm$ 0.003)\end{tabular}           & \begin{tabular}[c]{@{}c@{}}0.591 \\ ($\pm$ 0.003)\end{tabular}           & \begin{tabular}[c]{@{}c@{}}0.701 \\ ($\pm$ 0.005)\end{tabular}           & \begin{tabular}[c]{@{}c@{}}0.758 \\ ($\pm$ 0.004)\end{tabular}           & \begin{tabular}[c]{@{}c@{}}0.532 \\ ($\pm$ 0.002)\end{tabular}           & \begin{tabular}[c]{@{}c@{}}0.869 \\ ($\pm$ 0.001)\end{tabular}            & \begin{tabular}[c]{@{}c@{}}0.699 \\ ($\pm$ 0.003)\end{tabular}           & \begin{tabular}[c]{@{}c@{}}0.787 \\ ($\pm$ 0.003)\end{tabular}           \\
FRAME             & \begin{tabular}[c]{@{}c@{}}0.538 \\ ($\pm$ 0.004)\end{tabular}           & \begin{tabular}[c]{@{}c@{}}0.581 \\ ($\pm$ 0.004)\end{tabular}           & \begin{tabular}[c]{@{}c@{}}0.697 \\ ($\pm$ 0.004)\end{tabular}           & \begin{tabular}[c]{@{}c@{}}0.753 \\ ($\pm$ 0.004)\end{tabular}           & \begin{tabular}[c]{@{}c@{}}0.528 \\ ($\pm$ 0.002)\end{tabular}           & \begin{tabular}[c]{@{}c@{}}0.863 \\ ($\pm$ 0.001)\end{tabular}            & \begin{tabular}[c]{@{}c@{}}0.694 \\ ($\pm$ 0.003)\end{tabular}           & \begin{tabular}[c]{@{}c@{}}0.784 \\ ($\pm$ 0.003)\end{tabular}           \\
LightGT           & \begin{tabular}[c]{@{}c@{}}0.506 \\ ($\pm$ 0.003)\end{tabular}           & \begin{tabular}[c]{@{}c@{}}0.563 \\ ($\pm$ 0.003)\end{tabular}           & \begin{tabular}[c]{@{}c@{}}0.660 \\ ($\pm$ 0.004)\end{tabular}           & \begin{tabular}[c]{@{}c@{}}0.720 \\ ($\pm$ 0.004)\end{tabular}           & \begin{tabular}[c]{@{}c@{}}0.505 \\ ($\pm$ 0.001)\end{tabular}           & \begin{tabular}[c]{@{}c@{}}0.856 \\ ($\pm$ 0.001)\end{tabular}            & \begin{tabular}[c]{@{}c@{}}0.664 \\ ($\pm$ 0.003)\end{tabular}           & \begin{tabular}[c]{@{}c@{}}0.760 \\ ($\pm$ 0.003)\end{tabular}          \\
\hline
\end{tabular}
\caption{\label{tab:result1}The experimental results on the MVA dataset. The highest scores are marked in bold, with statistically significant paired \( t \)-test results (\( p=0.01 \)) indicated by an asterisk (*).}
\end{table*}

\begin{table*}[!t]
\centering
\begin{tabular}{c cccccccc}
\hline
Model & Precision@3 & Recall@3 & MAP@3 & NDCG@3 & Precision@5 & Recall@5 & MAP@5 & NDCG@5 \\ 
\hline
Proposed & \textbf{\begin{tabular}[c]{@{}c@{}}0.279*\\ ($\pm$ 0.004)\end{tabular}} & \textbf{\begin{tabular}[c]{@{}c@{}}0.632*\\ ($\pm$ 0.009)\end{tabular}} & \textbf{\begin{tabular}[c]{@{}c@{}}0.545* \\ ($\pm$ 0.010)\end{tabular}} & \textbf{\begin{tabular}[c]{@{}c@{}}0.591*\\ ($\pm$ 0.009)\end{tabular}} & \textbf{\begin{tabular}[c]{@{}c@{}}0.234*\\ ($\pm$ 0.002)\end{tabular}} & \textbf{\begin{tabular}[c]{@{}c@{}}0.760*\\ ($\pm$ 0.006)\end{tabular}} & \textbf{\begin{tabular}[c]{@{}c@{}}0.565*\\ ($\pm$ 0.008)\end{tabular}} & \textbf{\begin{tabular}[c]{@{}c@{}}0.637* \\ ($\pm$ 0.008)\end{tabular}} \\
BM3      & \begin{tabular}[c]{@{}c@{}}0.214 \\ ($\pm$ 0.014)\end{tabular}           & \begin{tabular}[c]{@{}c@{}}0.497 \\ ($\pm$ 0.029)\end{tabular}           & \begin{tabular}[c]{@{}c@{}}0.386 \\ ($\pm$ 0.037)\end{tabular}            & \begin{tabular}[c]{@{}c@{}}0.433 \\ ($\pm$ 0.037)\end{tabular}           & \begin{tabular}[c]{@{}c@{}}0.198 \\ ($\pm$ 0.009)\end{tabular}           & \begin{tabular}[c]{@{}c@{}}0.646 \\ ($\pm$ 0.027)\end{tabular}           & \begin{tabular}[c]{@{}c@{}}0.417 \\ ($\pm$ 0.035)\end{tabular}           & \begin{tabular}[c]{@{}c@{}}0.495 \\ ($\pm$ 0.035)\end{tabular}            \\
FRAME    & \begin{tabular}[c]{@{}c@{}}0.263 \\ ($\pm$ 0.006)\end{tabular}           & \begin{tabular}[c]{@{}c@{}}0.606 \\ ($\pm$ 0.013)\end{tabular}           & \begin{tabular}[c]{@{}c@{}}0.501 \\ ($\pm$ 0.015)\end{tabular}            & \begin{tabular}[c]{@{}c@{}}0.551 \\ ($\pm$ 0.015)\end{tabular}           & \begin{tabular}[c]{@{}c@{}}0.227 \\ ($\pm$ 0.003)\end{tabular}           & \begin{tabular}[c]{@{}c@{}}0.744 \\ ($\pm$ 0.009)\end{tabular}           & \begin{tabular}[c]{@{}c@{}}0.526 \\ ($\pm$ 0.014)\end{tabular}           & \begin{tabular}[c]{@{}c@{}}0.604 \\ ($\pm$ 0.013)\end{tabular}            \\
LightGT           & \begin{tabular}[c]{@{}c@{}}0.169 \\ ($\pm$ 0.002)\end{tabular}           & \begin{tabular}[c]{@{}c@{}}0.411 \\ ($\pm$ 0.006)\end{tabular}           & \begin{tabular}[c]{@{}c@{}}0.291 \\ ($\pm$ 0.003)\end{tabular}           & \begin{tabular}[c]{@{}c@{}}0.335 \\ ($\pm$ 0.004)\end{tabular}           & \begin{tabular}[c]{@{}c@{}}0.170 \\ ($\pm$ 0.001)\end{tabular}           & \begin{tabular}[c]{@{}c@{}}0.564 \\ ($\pm$ 0.005)\end{tabular}            & \begin{tabular}[c]{@{}c@{}}0.326 \\ ($\pm$ 0.003)\end{tabular}           & \begin{tabular}[c]{@{}c@{}}0.402 \\ ($\pm$ 0.003)\end{tabular}          \\
\hline
\end{tabular}
\caption{\label{tab:result2} The experimental results on the KuaiRand-Pure dataset. The highest scores are marked in bold, with statistically significant paired \( t \)-test results (\( p=0.01 \)) indicated by an asterisk (*).
}
\end{table*}

\subsubsection{BCE Loss for Supervised Learning.}

To further enhance the supervision, we integrate binary cross-entropy (BCE) loss based on whether a video was skipped or not. BCE loss treats the problem as a binary classification task, where the label \( y = 1 \) indicates that a video was not skipped, and \( y = 0 \) indicates that the video was skipped. The BCE loss is defined as
\begin{equation}
\mathcal{L}_{BCE} = - \left[ y \cdot \ln(\sigma_s(\hat{z})) + (1 - y) \cdot \ln(1 - \sigma_s(\hat{z})) \right],
\end{equation}
where \( y \) is the true binary label, and \( \hat{z} \) is the logit, which is the output before applying the sigmoid function to obtain the predicted probability that the video was not skipped. This loss enables the model to rank videos while simultaneously learning to classify them correctly based on whether they were skipped, providing a supervised learning signal that complements the ranking provided by the BPR loss.

\subsubsection{Combined Loss.}

The final combined loss function integrates the averaged BPR loss with the BCE loss, enabling the model to learn from both ranking and classification perspectives. The combined loss is given by
\begin{equation}
\mathcal{L}_{\text{combined}} = \lambda \mathcal{L}_{\text{BPR}} + (1 - \lambda) \mathcal{L}_{\text{BCE}},
\end{equation}
where \( \lambda \) is a hyperparameter that balances the contributions of the BPR loss and the BCE loss.

By combining these losses, the model benefits from the strengths of both approaches: the BPR loss ensures effective ranking of items according to nuanced user preferences, while the BCE loss supervises the model in accurately classifying videos based on user skip behavior. This comprehensive learning process results in a more refined and accurate recommendation system capable of both ranking items and predicting user engagement.

\section{Experiments}

\subsection{Experimental Settings}

\subsubsection{Datasets.} Table~\ref{tab:dataset_statistics} shows brief statistics of two datasets. The MVA dataset~\cite{10.1145/3539618.3591713} was collected from a mobile app platform. MVA includes interaction records containing user IDs, video IDs, playing time, video duration, interaction timestamps, and multi-level user behaviors such as likes, follows, and forwards. This dataset provides a rich context for analyzing user behavior by leveraging playing time and video duration to calculate the skip time for each user. For visual features, a pre-trained convolutional network is first used to extract features from each frame of all videos~\cite{10.1145/3539618.3591713}. Then, $K$ frames are sampled from each video, and features are extracted for each frame. These features are averaged to form a 128-dimensional vector. The MVA dataset consists of 12,739 users, 58,291 videos, and 342,694 interactions. The KuaiRand-Pure dataset~\cite{10.1145/3511808.3557624}, collected from the Kuaishou app, one of the largest video-sharing platforms in China, provides an unbiased sequential recommendation dataset. This dataset distinguishes itself by intervening in the recommendation policies of the platform through random insertion of selected videos over a two-week period, allowing for the collection of genuine user feedback without their awareness. The dataset captures 12 types of feedback signals, including clicks, favorites, and view time. The KuaiRand-Pure dataset includes 27,285 users, 7,583 videos, and 1,186,059 interactions, depending on the recommendation policy applied. Since the KuaiRand-Pure dataset does not provide video features, we set up learnable parameters for node embeddings. 

\subsubsection{Baselines.} We employed three state-of-the-art baselines in our experiments. Specifically, we used two multi-modal micro-video recommendation models, FRAME~\cite{10.1145/3539618.3591713} and LightGT~\cite{wei2023lightgt}, and one multi-media recommendation model, BM3 \cite{10.1145/3543507.3583251}. 
FRAME constructs a positive-negative graph for clip-level learning and models a dual-side GCN layer. 
LightGT inherits from the LightGCN model and Transformer, developing a modal-specific embedding and a layer-wise position encoder. 
BM3 bootstraps latent contrastive views in user-item representations, utilizing dropout augmentation.

\begin{figure}[t]
\centering
\includegraphics[width=0.9\columnwidth]{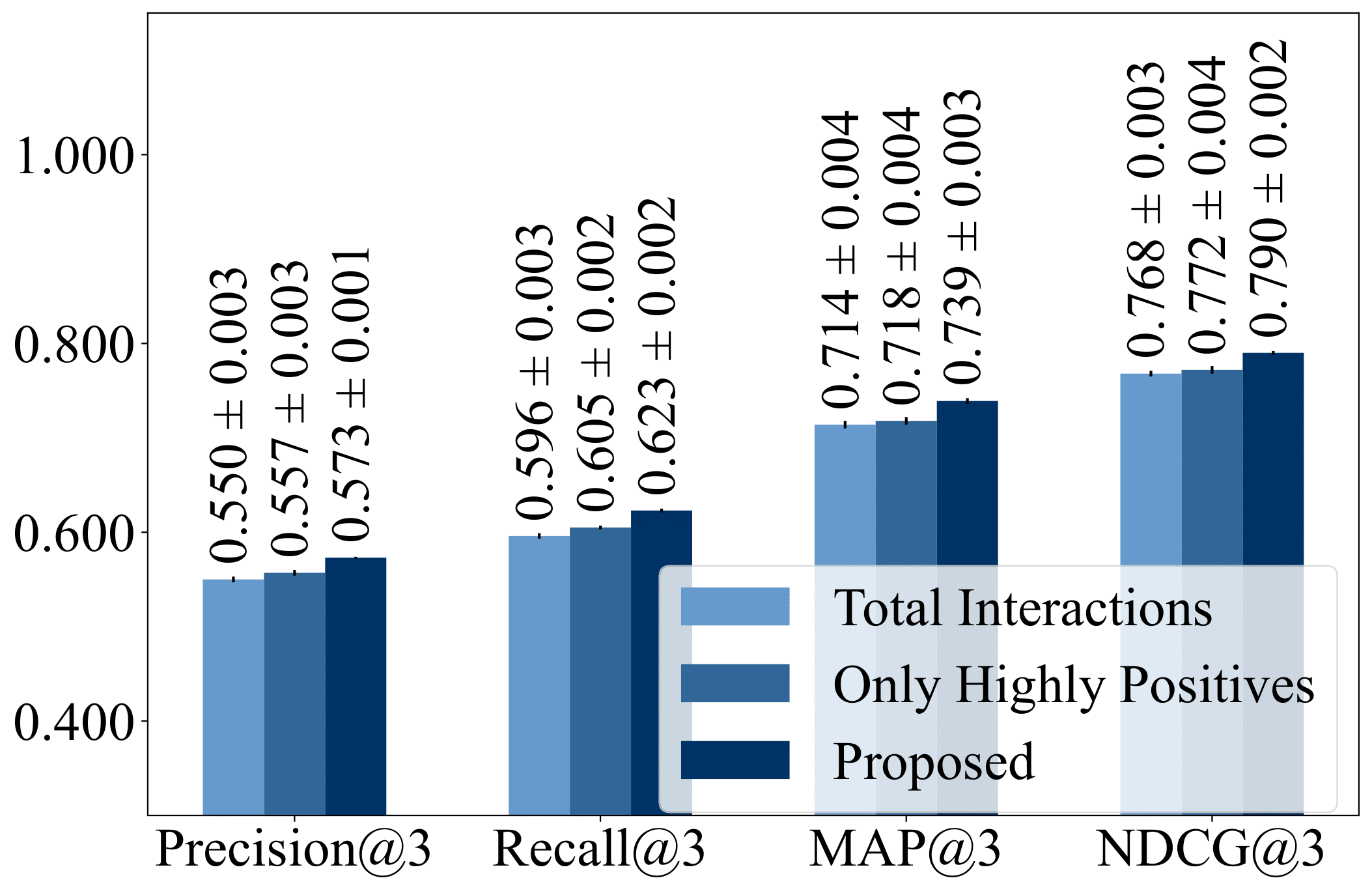}
\caption{ Comparison results between total interaction, highly positive only, and proposed dual-level graph. }
\label{fig:graph_ablation}
\end{figure}

\begin{figure}[t]
\centering
\includegraphics[width=0.9\columnwidth]{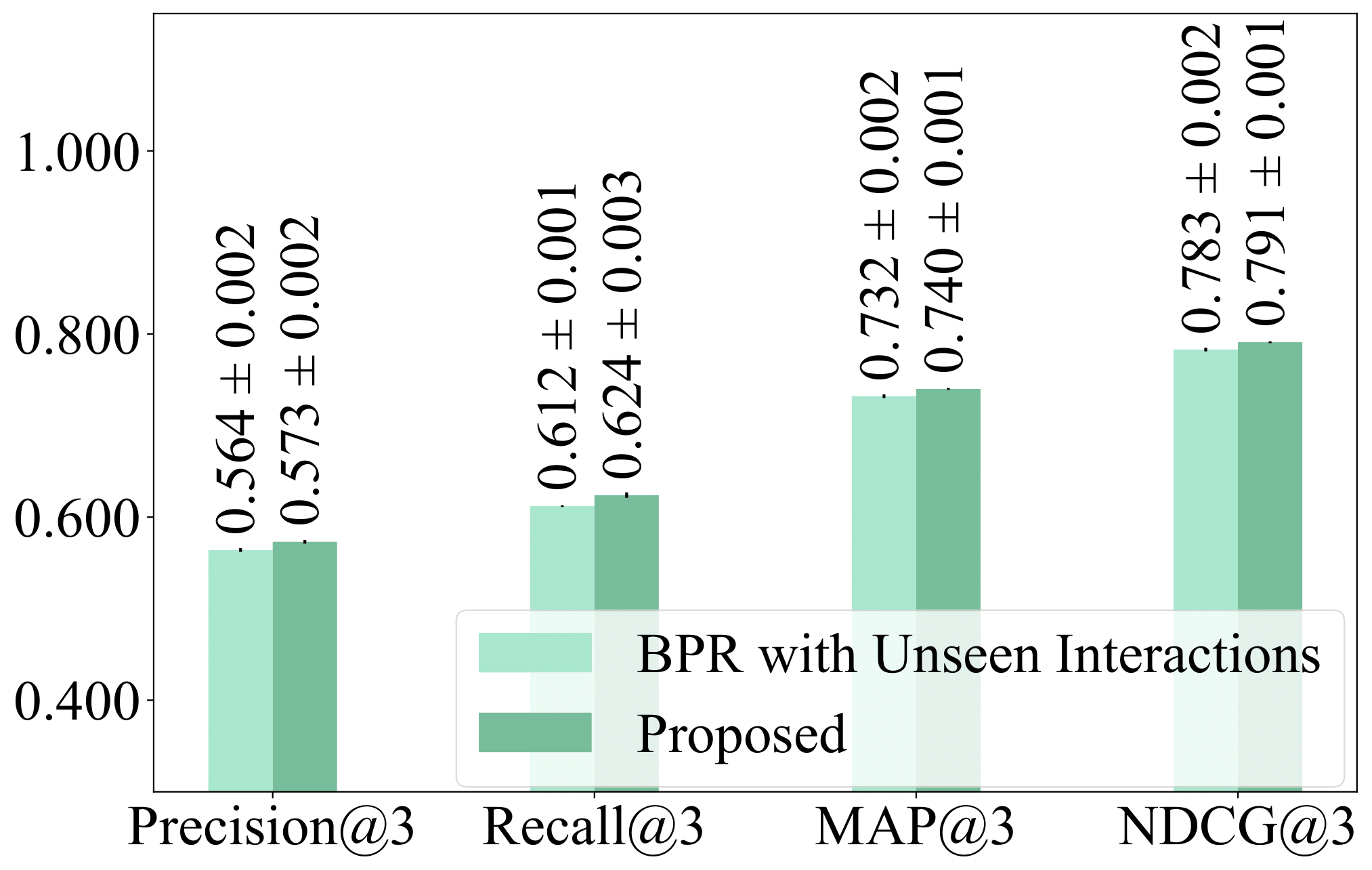}
\caption{ Comparison results between BPR loss with unseen interactions and proposed BRP loss.}
\label{fig:rank_ablation}
\end{figure}

\subsubsection{Implementation Details.} All models were implemented using PyTorch 2.3. The entire set of interactions is randomly sampled at a 6:2:2 ratio for each user to generate the training, validation, and test sets. Each model utilizes the training set to construct the adjacency matrix and is trained for up to 30 epochs. Early stopping is applied if the recall@3 on the validation set does not improve for five consecutive epochs. The evaluation measures include Top-$k$ recall, precision, mean average precision (MAP), and normalized discounted cumulative gain (NDCG) at $k = \{3, 5\}$. Preference prediction is conducted on the items associated with users in the test interactions.

To ensure robust results in a statistical way, data splitting, training, and testing were repeated ten times in the MVA dataset and seven times in the KuaiRand-Pure dataset. For performance comparison, statistical significance was assessed using a paired $t$-test implemented via the SciPy open-source library. Each model was trained with a batch size of 1024 using the AdamW optimizer \cite{RN16} with a momentum of 0.9 and a weight decay of 1e-4. The learning rate starts at 1e-3 and decays to 1e-6 following a cosine annealing schedule. Each model was trained based on the loss function suggested in the original papers. The feature dimension $d$ is set to 128, and $\lambda$ is set to 0.5.

\subsection{Experimental Results}

\subsubsection{Comparison Results.} 

As shown in Tables~\ref{tab:result1} and~\ref{tab:result2}, the two experimental results demonstrate that the proposed method outperforms the comparison models across two different datasets. Table~\ref{tab:result1} presents results on the MVA dataset, the proposed method achieved the highest scores in all evaluation measures. Notably, the proposed method significantly outperformed other models with statistical significance at the p=0.01 level across all measures. Table~\ref{tab:result2} shows the results on the KuaiRand-Pure dataset, where the proposed method again achieved the best performance in all evaluation measures. In this dataset, the proposed method also significantly outperformed the other models with a p=0.01 level of statistical significance. Both experiments consistently demonstrate the superior performance of the proposed method compared to the conventional models. This consistent superiority across different datasets highlights the efficacy of the proposed method. 

\subsubsection{Ablation Study.}

Figures~\ref{fig:graph_ablation} and~\ref{fig:rank_ablation} illustrate the experimental results of the ablation study on the proposed model. As shown in Figure \ref{fig:graph_ablation}, we examined the performance of different graph construction methods. In the field of micro-video recommendation, the model can be trained using all interactions as positive signals, especially when there is a lack of explicit feedback or playing time data. Alternatively, playing time can be used to filter out data where skipping occurs, allowing the model to learn high-confidence user preferences. The proposed dual-level positive graph construction method outperformed both of these approaches. Using total interaction data without distinguishing is overly naive and can hinder the learning process. Furthermore, the relatively superior performance of the proposed model compared to the only highly approach suggests that less positive interactions can provide valuable information for interaction modeling.

As seen in Figure~\ref{fig:rank_ablation}, we compared the widely used negative sampling for unseen interactions with the proposed BPR loss. The proposed BPR loss demonstrated superior performance, indicating that negative interactions that mean quick skip have a clearly defined relative ranking compared to highly positive interactions that were fully viewed. Furthermore, the dual BPR loss, which includes the less positive interactions, constructs a hierarchical ranking, suggesting that it warrants further investigation in future research.

\section{Conclusion}

This study proposes a dual-graph-based micro-video recommender system that effectively utilizes the granular details of user interactions, particularly by distinguishing based on skip behaviors between fully-viewed interactions, delayed skips, and quick skips. The experimental results demonstrate that our approach outperforms three conventional methods across eight evaluation measures on two public micro-video datasets. 

In future research, we aim to develop methods that are not dependent on specific thresholds, such as 5 seconds, enabling their application to a wide range of datasets. For instance, in the KuaiRec dataset~\cite{10.1145/3511808.3557220}, many skips occur even after 10 seconds. This indicates that user behavior may vary depending on the platform. In addition, advanced designs such as attention mechanisms~\cite{tao2020mgat,liu2022multi} can be incorporated into submodules to enhance the efficacy of the proposed model.

\section{Acknowledgments}
This research was supported in part by the Institute of Information \& Communications Technology Planning \& Evaluation (IITP) grant funded by the Korean Government (MSIT) (2021-0-01341, Artificial Intelligence Graduate School Program (Chung-Ang University)), and in part by the National Research Foundation of Korea (NRF) grant funded by the Korea government (MSIT) (No. RS-2024-00459387).
We are grateful to the reviewers for their insightful suggestions during the rebuttal process, which enhanced the quality of this paper and provided guidance for future studies.
% \bibliography{main_ref}

\begin{thebibliography}{44}
\providecommand{\natexlab}[1]{#1}

\bibitem[{Banerjee and Pal(2021)}]{9377378}
Banerjee, S.; and Pal, A. 2021.
\newblock Skipping Skippable Ads on YouTube: How, When, Why and Why Not?
\newblock In \emph{Proceedings of 2021 15th International Conference on Ubiquitous Information Management and Communication}, 1--5.

\bibitem[{Cai et~al.(2022{\natexlab{a}})Cai, Qian, Fang, Hu, Ding, and Xu}]{cai2022heterogeneous}
Cai, D.; Qian, S.; Fang, Q.; Hu, J.; Ding, W.; and Xu, C. 2022{\natexlab{a}}.
\newblock Heterogeneous Graph Contrastive Learning Network for Personalized Micro-Video Recommendation.
\newblock \emph{IEEE Transactions on Multimedia}, 25: 2761--2773.

\bibitem[{Cai et~al.(2022{\natexlab{b}})Cai, Qian, Fang, Hu, and Xu}]{cai2022adaptive}
Cai, D.; Qian, S.; Fang, Q.; Hu, J.; and Xu, C. 2022{\natexlab{b}}.
\newblock Adaptive Anti-Bottleneck Multi-Modal Graph Learning Network for Personalized Micro-Video Recommendation.
\newblock In \emph{Proceedings of the 30th ACM International Conference on Multimedia}, 581--590.

\bibitem[{Cai et~al.(2021)Cai, Qian, Fang, and Xu}]{cai2021heterogeneous}
Cai, D.; Qian, S.; Fang, Q.; and Xu, C. 2021.
\newblock Heterogeneous Hierarchical Feature Aggregation Network for Personalized Micro-Video Recommendation.
\newblock \emph{IEEE Transactions on Multimedia}, 24: 805--818.

\bibitem[{{Chaves}(2024)}]{MV1}
{Chaves}. 2024.
\newblock 20+ Interesting Short Form Video Statistics \& Trends (2024).
\newblock \url{https://vidico.com/news/short-form-video-statistics}.
\newblock Accessed: 2024-08-14.

\bibitem[{Chen et~al.(2018)Chen, Liu, Zha, Zhou, Xiong, and Li}]{chen2018temporal}
Chen, X.; Liu, D.; Zha, Z.-J.; Zhou, W.; Xiong, Z.; and Li, Y. 2018.
\newblock Temporal Hierarchical Attention at Category- And Item-Level for Micro-Video Click-Through Prediction.
\newblock In \emph{Proceedings of the 26th ACM International Conference on Multimedia}, 1146--1153.

\bibitem[{Du et~al.(2022)Du, Wu, Feng, He, and Tang}]{du2022invariant}
Du, X.; Wu, Z.; Feng, F.; He, X.; and Tang, J. 2022.
\newblock Invariant Representation Learning for Multimedia Recommendation.
\newblock In \emph{Proceedings of the 30th ACM International Conference on Multimedia}, 619--628.

\bibitem[{Gao et~al.(2022{\natexlab{a}})Gao, Li, Lei, Chen, Li, Jiang, He, Mao, and Chua}]{10.1145/3511808.3557220}
Gao, C.; Li, S.; Lei, W.; Chen, J.; Li, B.; Jiang, P.; He, X.; Mao, J.; and Chua, T.-S. 2022{\natexlab{a}}.
\newblock KuaiRec: A Fully-Observed Dataset and Insights for Evaluating Recommender Systems.
\newblock In \emph{Proceedings of the 31st ACM International Conference on Information \& Knowledge Management}, 540–550.

\bibitem[{Gao et~al.(2022{\natexlab{b}})Gao, Li, Zhang, Chen, Li, Lei, Jiang, and He}]{10.1145/3511808.3557624}
Gao, C.; Li, S.; Zhang, Y.; Chen, J.; Li, B.; Lei, W.; Jiang, P.; and He, X. 2022{\natexlab{b}}.
\newblock KuaiRand: An Unbiased Sequential Recommendation Dataset With Randomly Exposed Videos.
\newblock In \emph{Proceedings of the 31st ACM International Conference on Information \& Knowledge Management}, 3953–3957.

\bibitem[{Gu and Hu(2024)}]{GU2024111299}
Gu, P.; and Hu, H. 2024.
\newblock A Holistic View on Positive and Negative Implicit Feedback for Micro-Video Recommendation.
\newblock \emph{Knowledge-Based Systems}, 284: 111299.

\bibitem[{He et~al.(2020)He, Deng, Wang, Li, Zhang, and Wang}]{10.1145/3397271.3401063}
He, X.; Deng, K.; Wang, X.; Li, Y.; Zhang, Y.; and Wang, M. 2020.
\newblock LightGCN: Simplifying and Powering Graph Convolution Network for Recommendation.
\newblock In \emph{Proceedings of the 43rd International ACM SIGIR Conference on Research and Development in Information Retrieval}, 639–648.

\bibitem[{Jia et~al.(2022)Jia, Zhou, Chen, and Shi}]{10.1007/978-3-031-05897-4_29}
Jia, W.; Zhou, R.; Chen, N.; and Shi, Y. 2022.
\newblock Examining the Usability of a Short-Video App Interface Through an Eye-Tracking Experiment.
\newblock In Soares, M.~M.; Rosenzweig, E.; and Marcus, A., eds., \emph{Design, User Experience, and Usability: UX Research, Design, and Assessment}, 414--427. Cham.

\bibitem[{Jiang et~al.(2020)Jiang, Wang, Wei, Gao, Wang, and Nie}]{10.1145/3394171.3413653}
Jiang, H.; Wang, W.; Wei, Y.; Gao, Z.; Wang, Y.; and Nie, L. 2020.
\newblock What Aspect Do You Like: Multi-Scale Time-Aware User Interest Modeling for Micro-Video Recommendation.
\newblock In \emph{Proceedings of the 28th ACM International Conference on Multimedia}, 3487–3495.

\bibitem[{{Kelemen}(2023)}]{MV4}
{Kelemen}. 2023.
\newblock Reach More People by Optimizing the First 5 Seconds of Your Video.
\newblock \url{https://www.wintersummer.ca/blog/the-first-five-seconds-of-your-video-are-most-important}.
\newblock Accessed: 2024-08-14.

\bibitem[{Lei et~al.(2023)Lei, Cao, Yang, Ding, and Zhang}]{lei2023learning}
Lei, F.; Cao, Z.; Yang, Y.; Ding, Y.; and Zhang, C. 2023.
\newblock Learning the User’s Deeper Preferences for Multi-Modal Recommendation Systems.
\newblock \emph{ACM Transactions on Multimedia Computing, Communications and Applications}, 19(3s): 1--18.

\bibitem[{Li et~al.(2019)Li, Liu, Yin, Cui, Xu, and Nie}]{li2019routing}
Li, Y.; Liu, M.; Yin, J.; Cui, C.; Xu, X.-S.; and Nie, L. 2019.
\newblock Routing Micro-Videos via a Temporal Graph-Guided Recommendation System.
\newblock In \emph{Proceedings of the 27th ACM International Conference on Multimedia}, 1464--1472.

\bibitem[{Liu, Li, and Tian(2022)}]{liu2022multi}
Liu, H.; Li, C.; and Tian, L. 2022.
\newblock Multi-Modal Graph Attention Network for Video Recommendation.
\newblock In \emph{2022 IEEE International Conference on Computer and Communication Engineering Technology}, 94--99.

\bibitem[{Liu et~al.(2019)Liu, Chen, Liu, and Hu}]{liu2019user}
Liu, S.; Chen, Z.; Liu, H.; and Hu, X. 2019.
\newblock User-Video Co-Attention Network for Personalized Micro-Video Recommendation.
\newblock In \emph{Proceedings of the World Wide Web Conference}, 3020--3026.

\bibitem[{Liu et~al.(2020)Liu, Xie, Zou, and Chen}]{liu2020user}
Liu, S.; Xie, J.; Zou, C.; and Chen, Z. 2020.
\newblock User Conditional Hashtag Recommendation for Micro-Videos.
\newblock In \emph{Proceedings of the 2020 IEEE International Conference on Multimedia and Expo}, 1--6.

\bibitem[{Liu et~al.(2022)Liu, Tao, Shao, Yang, and Huang}]{liu2022elimrec}
Liu, X.; Tao, Z.; Shao, J.; Yang, L.; and Huang, X. 2022.
\newblock EliMRec: Eliminating Single-Modal Bias in Multimedia Recommendation.
\newblock In \emph{Proceedings of the 30th ACM International Conference on Multimedia}, 687--695.

\bibitem[{Liu et~al.(2021)Liu, Liu, Tian, Wang, Niu, Song, and Li}]{liu2021concept}
Liu, Y.; Liu, Q.; Tian, Y.; Wang, C.; Niu, Y.; Song, Y.; and Li, C. 2021.
\newblock Concept-Aware Denoising Graph Neural Network for Micro-Video Recommendation.
\newblock In \emph{Proceedings of the 30th ACM International Conference on Information \& Knowledge Management}, 1099--1108.

\bibitem[{Loshchilov and Hutter(2019)}]{RN16}
Loshchilov, I.; and Hutter, F. 2019.
\newblock Decoupled Weight Decay Regularization.
\newblock In \emph{Proceddings of the International Conference on Learning Representation}, 1--18.

\bibitem[{Lu et~al.(2023)Lu, Huang, Zhang, Han, Chen, Fan, Lai, Zhao, and Wu}]{lu2023multi}
Lu, Y.; Huang, Y.; Zhang, S.; Han, W.; Chen, H.; Fan, W.; Lai, J.; Zhao, Z.; and Wu, F. 2023.
\newblock Multi-Trends Enhanced Dynamic Micro-Video Recommendation.
\newblock In \emph{Proceedings of the International Conference on Artificial Intelligence}, 430--441.

\bibitem[{{Park}(2023)}]{MV3}
{Park}. 2023.
\newblock Social Media/Search Portal Trend Report.
\newblock \url{ https://blog.opensurvey.co.kr/trendreport/socialmedia-2023}.
\newblock Accessed: 2024-08-14.

\bibitem[{{Perez}(2024)}]{MV2}
{Perez}. 2024.
\newblock YouTube Says Over 25% of Its Creator Partners Now Monetize via Shorts.
\newblock \url{https://techcrunch.com/2024/03/28/youtube-says-over-25-of-its-creator-partners-now-monetize-via-shorts}.
\newblock Accessed: 2024-08-14.

\bibitem[{Rendle et~al.(2020)Rendle, Krichene, Zhang, and Anderson}]{rendle2020neural}
Rendle, S.; Krichene, W.; Zhang, L.; and Anderson, J. 2020.
\newblock Neural Collaborative Filtering vs. Matrix Factorization Revisited.
\newblock In \emph{Proceedings of the 14th ACM Conference on Recommender Systems}, 240--248.

\bibitem[{Shang et~al.(2023)Shang, Gao, Chen, Jin, Wang, and Li}]{10.1145/3539618.3591713}
Shang, Y.; Gao, C.; Chen, J.; Jin, D.; Wang, M.; and Li, Y. 2023.
\newblock Learning Fine-Grained User Interests for Micro-Video Recommendation.
\newblock In \emph{Proceedings of the 46th International ACM SIGIR Conference on Research and Development in Information Retrieval}, 433–442.

\bibitem[{Tao et~al.(2022)Tao, Liu, Xia, Wang, Yang, Huang, and Chua}]{tao2022self}
Tao, Z.; Liu, X.; Xia, Y.; Wang, X.; Yang, L.; Huang, X.; and Chua, T.-S. 2022.
\newblock Self-Supervised Learning for Multimedia Recommendation.
\newblock \emph{IEEE Transactions on Multimedia}, 25: 5107--5116.

\bibitem[{Tao et~al.(2020)Tao, Wei, Wang, He, Huang, and Chua}]{tao2020mgat}
Tao, Z.; Wei, Y.; Wang, X.; He, X.; Huang, X.; and Chua, T.-S. 2020.
\newblock Mgat: Multimodal Graph Attention Network for Recommendation.
\newblock \emph{Information Processing \& Management}, 57(5): 102277.

\bibitem[{Tian et~al.(2022)Tian, Chang, Niu, Song, and Li}]{10.1145/3477495.3532081}
Tian, Y.; Chang, J.; Niu, Y.; Song, Y.; and Li, C. 2022.
\newblock When Multi-Level Meets Multi-Interest: A Multi-Grained Neural Model for Sequential Recommendation.
\newblock In \emph{Proceedings of the 45th International ACM SIGIR Conference on Research and Development in Information Retrieval}, 1632–1641.

\bibitem[{Wang et~al.(2021)Wang, Wei, Yin, Wu, Song, and Nie}]{wang2021dualgnn}
Wang, Q.; Wei, Y.; Yin, J.; Wu, J.; Song, X.; and Nie, L. 2021.
\newblock Dualgnn: Dual Graph Neural Network for Multimedia Recommendation.
\newblock \emph{IEEE Transactions on Multimedia}, 25: 1074--1084.

\bibitem[{Wang et~al.(2019)Wang, He, Wang, Feng, and Chua}]{wang2019neural}
Wang, X.; He, X.; Wang, M.; Feng, F.; and Chua, T.-S. 2019.
\newblock Neural Graph Collaborative Filtering.
\newblock In \emph{Proceedings of the 42nd International ACM SIGIR Conference on Research and Development in Information Retrieval}, 165--174.

\bibitem[{Wang et~al.(2020)Wang, Jin, Zhang, He, Xu, and Chua}]{wang2020disentangled}
Wang, X.; Jin, H.; Zhang, A.; He, X.; Xu, T.; and Chua, T.-S. 2020.
\newblock Disentangled Graph Collaborative Filtering.
\newblock In \emph{Proceedings of the 43rd International ACM SIGIR Conference on Research and Development in Information Retrieval}, 1001--1010.

\bibitem[{Wang(2021)}]{Wang_2021}
Wang, Y. 2021.
\newblock Content Characteristics and Limitations of Original Short Video Based on Depth Data.
\newblock \emph{Journal of Physics: Conference Series}, 1881(4): 042070.

\bibitem[{Wang, Wu, and Hoashi(2019)}]{10.1145/3340555.3355720}
Wang, Y.; Wu, J.; and Hoashi, K. 2019.
\newblock Multi-Attention Fusion Network for Video-Based Emotion Recognition.
\newblock In \emph{2019 International Conference on Multimodal Interaction}, 595–601.

\bibitem[{Wei et~al.(2023{\natexlab{a}})Wei, Huang, Xia, and Zhang}]{wei2023multi}
Wei, W.; Huang, C.; Xia, L.; and Zhang, C. 2023{\natexlab{a}}.
\newblock Multi-Modal Self-Supervised Learning for Recommendation.
\newblock In \emph{Proceedings of the ACM Web Conference 2023}, 790--800.

\bibitem[{Wei et~al.(2023{\natexlab{b}})Wei, Liu, Liu, Wang, Nie, and Chua}]{wei2023lightgt}
Wei, Y.; Liu, W.; Liu, F.; Wang, X.; Nie, L.; and Chua, T.-S. 2023{\natexlab{b}}.
\newblock {LightGT}: A Light Graph Transformer for Multimedia Recommendation.
\newblock In \emph{Proceedings of the 46th International ACM SIGIR Conference on Research and Development in Information Retrieval}, 1508--1517.

\bibitem[{Wei et~al.(2021)Wei, Wang, He, Nie, Rui, and Chua}]{wei2021hierarchical}
Wei, Y.; Wang, X.; He, X.; Nie, L.; Rui, Y.; and Chua, T.-S. 2021.
\newblock Hierarchical User Intent Graph Network for Multimedia Recommendation.
\newblock \emph{IEEE Transactions on Multimedia}, 24: 2701--2712.

\bibitem[{Wei et~al.(2020)Wei, Wang, Nie, He, and Chua}]{wei2020graph}
Wei, Y.; Wang, X.; Nie, L.; He, X.; and Chua, T.-S. 2020.
\newblock Graph-Refined Convolutional Network for Multimedia Recommendation With Implicit Feedback.
\newblock In \emph{Proceedings of the 28th ACM International Conference on Multimedia}, 3541--3549.

\bibitem[{Wei et~al.(2019)Wei, Wang, Nie, He, Hong, and Chua}]{10.1145/3343031.3351034}
Wei, Y.; Wang, X.; Nie, L.; He, X.; Hong, R.; and Chua, T.-S. 2019.
\newblock MMGCN: Multi-Modal Graph Convolution Network for Personalized Recommendation of Micro-Video.
\newblock In \emph{Proceedings of the 27th ACM International Conference on Multimedia}, 1437–1445.

\bibitem[{Willis(2024)}]{Willis2024}
Willis, M. 2024.
\newblock \emph{Short Video Adverts: A Modern and Virtual Form of Advertising}, 107--131.
\newblock Cham: Springer International Publishing.

\bibitem[{Yi et~al.(2022)Yi, Wang, Ounis, and Macdonald}]{yi2022multi}
Yi, Z.; Wang, X.; Ounis, I.; and Macdonald, C. 2022.
\newblock Multi-Modal Graph Contrastive Learning for Micro-Video Recommendation.
\newblock In \emph{Proceedings of the 45th International ACM SIGIR Conference on Research and Development in Information Retrieval}, 1807--1811.

\bibitem[{Zhang, Wang, and Ariffin(2024)}]{ZHANG2024104014}
Zhang, Q.; Wang, Y.; and Ariffin, S.~K. 2024.
\newblock Keep Scrolling: An Investigation of Short Video Users’ Continuous Watching Behavior.
\newblock \emph{Information \& Management}, 61(6): 104014.

\bibitem[{Zhou et~al.(2023)Zhou, Zhou, Liu, Zeng, Miao, Wang, You, and Jiang}]{10.1145/3543507.3583251}
Zhou, X.; Zhou, H.; Liu, Y.; Zeng, Z.; Miao, C.; Wang, P.; You, Y.; and Jiang, F. 2023.
\newblock Bootstrap Latent Representations for Multi-Modal Recommendation.
\newblock In \emph{Proceedings of the ACM Web Conference 2023}, 845–854.

\end{thebibliography}

\end{document}